\documentclass[preprint,aps,prd,tightenlines,nofootinbib,superscriptaddress]{revtex4-1}

\usepackage{color}

\begin{document}

\title{ LIGHT-FRONT QUANTUM CHROMODYNAMICS \\
A framework for the analysis of hadron physics}

\author{B.L.G. Bakker (VU Amsterdam)}
\altaffiliation{Member of White Paper Development Committee.}
\author{A. Bassetto (INFN-Padova)}
\author{S. J. Brodsky (SLAC, Stanford U)}
\author{W. Broniowski (Jan Kochanowski U)}
\author{S. Dalley (SMU)}
\author{T. Frederico (Inst Tecnologico de Aeronautica)}
\altaffiliation{Member of White Paper Development Committee.}
\author{S. D. G{\l}azek (U Warsaw)}
\altaffiliation{Member of White Paper Development Committee.}
\author{J. R. Hiller (U Minn-Duluth)}
\altaffiliation{Member of White Paper Development Committee.}
\author{C.-R. Ji (NCSU)}
\author{V. Karmanov (Lebedev Physical Inst)}
\author{D. Kulshreshtha (U Delhi)}
\author{J.-F. Mathiot (U Blaise Pascal)}
\author{W. Melnitchouk (Jefferson Lab)}
\author{G. A. Miller (U Washington)}
\author{J. Papavassiliou (U Valencia)}
\author{W. N. Polyzou (U Iowa)}
\author{N. G. Stefanis (Ruhr U Bochum)}
\author{J. P. Vary (Iowa State)}
\affiliation{Member of ILCAC, Inc.}
\author{A. Ilderton (Chalmers)}
\altaffiliation{Author of section on intense
time-dependent fields.}
%
\author{T. Heinzl (Plymouth)}
\altaffiliation{Author of section on intense
time-dependent fields.}

\date{ 17 September 2013 }

\preprint{SLAC-PUB-15745, JLAB-THY-13-1804}

\begin{abstract}

An outstanding goal of physics is to find
solutions that describe hadrons in the theory of
strong interactions, Quantum Chromodynamics (QCD).
For this goal, the light-front Hamiltonian
formulation of QCD (LFQCD) is a complementary
approach to the well-established lattice gauge
method. LFQCD offers access to the hadrons'
nonperturbative quark and gluon amplitudes, which
are directly testable in experiments at existing and future
facilities. We present an overview of the promises
and challenges of LFQCD in the context of unsolved
issues in QCD that require broadened and
accelerated investigation. 
We identify specific goals of this approach and address its quantifiable uncertainties.

\end{abstract}

\maketitle

\section{Introduction}

Quantum Chromodynamics (QCD), the theory of strong
interactions, is a part of the Standard Model of
elementary particles that also includes, besides
QCD, the theory of electro-weak (EW) interactions.
In view of the difference in strength of these
interactions, one may treat the EW interactions as
a perturbation in systems consisting of hadrons,
the composite particles that respond to the strong
interactions. Perturbation theory has its place in
QCD also, but only at large values of the
transferred energy or momentum where it exhibits
the property of asymptotic freedom. The field of
perturbative QCD is well developed and many
phenomena have been described using it, such as
factorization, parton distributions, single-spin
asymmetries, and jets. However, at low values of
the energy and momentum transfer, the strong
interaction must be treated in a nonperturbative
manner, since the interaction strength becomes
large and the confinement of quarks and gluons, as
the partonic components of the hadrons, cannot be
ignored. There is a wealth of data in this strong
interaction regime that is waiting for explanation
in terms of calculations proceeding directly from
the underlying theory. As one prominent
application of an {\em ab initio} approach to QCD,
we mention that many extensive experimental
programs either measure directly, or depend upon
the knowledge of, the probability distributions of
the quark and gluon components of the hadrons.

Three approaches have produced considerable
success in the strong-coupling area up to the
present. First, hadronic models have been
formulated and applied successfully, such as in
Refs.~\cite{m1,m2,m3,m4, m5,m6,m7,m8}. This
success comes sometimes at the price of
introducing parameters that need to be identified
quantitatively. For example, the Relativistic
String Hamiltonian developed by Simonov et al.
\cite{S1,S2,S3,S4,S5,S6,S7,S8} depends on the
current quark masses, the string tension, and a
parameter corresponding to $\Lambda_{\rm QCD}$.
The second method, lattice
QCD~\cite{L1,L2,L3,L4,L5}, is an {\em ab initio}
approach directly linked to the Lagrangian of QCD.
Based on a Euclidean formulation, lattice QCD
provides an estimate of the QCD path integral and
opens access to low-energy hadronic properties
such as masses. Although lattice QCD can estimate
some observables directly, it does not provide the
wave functions (WF) that are needed for the
description of the structure and dynamics of
hadrons. Third is the Dyson--Schwinger
approach~\cite{DS1,DS2,DS3}. It is also formulated
in Euclidean space-time and employs models for
vertex functions.

Light-Front QCD (LFQCD) is an alternative {\em ab
initio} approach to strongly interacting
systems~\cite{LFR1,LFR2,LFR3,LFR4,LFR5,LQ1,LQ2,
L3,LQ4,LQ5,LQ6,LQ7,LQ8,LQ9,LQ10,LQ11,
LQ12,LQ13,LQ14,DQ1,DQ2,DQ3,DQ4,DQ5,DQ6}. It is,
like perturbative and lattice QCD, directly
connected to the QCD Lagrangian, but it is a
Hamiltonian method, formulated in Minkowski space
rather than Euclidean space. The essential
ingredient is Dirac's front form of Hamiltonian
dynamics~\cite{Dirac1,Dirac2,Dirac3}, where one
quantizes the theory at fixed light-cone time
$\tau= t + z/c$ rather than ordinary time $t$. An
interpolation between the instant form and the
front form of the relativistic Hamiltonian
dynamics is discussed in Ref.~\cite{I1,I2,I3}.
Thus, initial conditions for a WF are set not at a
single time $t$, but on the space-time hyperplane
swept by the front of a plane wave of light. The
solutions will be exact mass spectra and
light-front wave functions (LFWFs) capable of
describing a wide range of experiments in a
relativistically covariant manner. For example,
one obtains the probability distributions of the
quark and gluon components of the hadrons from the
squared modulae of the LFWFs. Hence, LFQCD
exhibits the promise of accessing a much wider
range of experimental situations than previously
addressed.

The light-front framework has many attractive
features. On the technical side, LFQCD provides
the largest number of kinematic
(interaction-independent) generators of the
Poincar\'{e} transformations in relativistic
Hamiltonian dynamics, i.e., seven instead of only
six in other frameworks. The eigenvalues of the
LFQCD Hamiltonian are the discrete masses and
continuous invariant-mass hadronic spectra,
instead of the frame-dependent energies. The
method yields the boost-invariant and
process-independent LFWFs needed for form factors,
scattering amplitudes, correlations, spin effects,
decay rates, momentum space distributions, and
other hadronic observables. 

Quantization in the light-front provides the
field-theoretical realization of the intuitive
ideas of the parton model~\cite{Feynman:1969ej,
Feynman:1973xc} which is formulated at fixed $t$
in the infinite-momentum
frame~\cite{Fubini:1965xx, Weinberg:1966jm}. The
same results are obtained in the front form for
any frame; e.g., as already mentioned above, the
structure functions and other probabilistic parton
distributions measured in deep inelastic
scattering are obtained from the squares of the
boost invariant LFWFs, the eigensolution of the
light-front Hamiltonian. In particular, the
``handbag'' contributions~\cite{Brodsky:2000xy} to
the $E$ and $H$ generalized parton distributions
for deeply virtual Compton scattering, which can be
computed from the overlap of LFWFs, automatically
satisfy the known sum rules. The LFWFs contain
information about novel QCD features, such as
color transparency~\cite{ct1,ct2}, hidden
color~\cite{hc1,hc2,hc3,hc4,hc5,hc6,hc7},
intrinsic charm~\cite{ic1,ic2,ic3,ic4,ic5}, sea-quark
symmetries~\cite{sq1,sq2}, dijet
diffraction~\cite{dijetdiffraction}, direct hard
processes,~\cite{directhardprocesses}, and
hadronic spin dynamics~\cite{spin1,spin2,spin3}.
The familiar kinematic variable $x_{Bj}$ of deep
inelastic lepton-hadron scattering becomes
identified with the LF $+$-momentum fraction $x$
carried by the constituent in a hadron that is
struck by the gauge boson emitted by the lepton.
The BFKL Regge behavior of structure functions can
be demonstrated~\cite{Mueller:1993rr} from the
behavior of LFWFs at small $x$. Hadronic matrix
elements of currents can be obtained as overlaps
of LFWFs as in the Drell-Yan-West formula~\cite{
Drell:1969km,West:1970av,Brodsky:1980zm}. The 
gauge-invariant meson and baryon
distribution amplitudes which control hard
exclusive and direct reactions are the valence
LFWFs integrated over transverse momentum at fixed
$x$. The ``ERBL''
evolution~\cite{LQ7,Efremov:1979qk} of
distribution amplitudes and the factorization
theorems for hard exclusive processes can be
derived most directly using LF methods. 

One can also prove fundamental theorems for
relativistic quantum field theories using the
front form, including the cluster decomposition
theorem~\cite{Brodsky:1985gs} and the vanishing of
the anomalous gravitomagnetic moment for any Fock
state of a hadron~\cite{spin3}. One can
show that a nonzero anomalous magnetic moment of a
bound state requires nonzero angular momentum of
the constituents. The cluster
properties~\cite{Antonuccio:1997tw} of LF
time-ordered perturbation theory, together with
$J^z$ conservation, can be used to derive the
Parke-Taylor rules for multi-gluon scattering
amplitudes~\cite{Cruz-Santiago:2013vta}. The
counting-rule~\cite{Brodsky:1994kg, Farrar-Jackson, MEK} behavior of
structure functions at large $x$ and Bloom-Gilman
duality have also been derived in LFQCD. The
existence of ``lensing effects" at leading twist,
such as the $T$-odd ``Sivers effect" in
spin-dependent semi-inclusive deep-inelastic
scattering, was first demonstrated using LF
methods~\cite{Brodsky:2002cx}. 

LF quantization is thus the natural framework for
description of the nonperturbative relativistic
bound-state structure of hadrons using QCD.
However, there exist subtle problems in LFQCD that
require thorough investigation. For example, the
complexities of the vacuum in the usual
instant-time formulation~\cite{V1,V2,V3,V4,V5,L1,
V7,V8,V9,V10,V11,V12,V13,V14,V15,V16,V17,V18,LQ14,
V20}, such as the Higgs mechanism and condensates
in $\phi^4$ theory, have their counterparts in
zero modes or, possibly, in additional terms in
the LFQCD Hamiltonian that are allowed by power
counting~\cite{LQ14}. LF considerations of the
vacuum as well as the problem of achieving full
covariance in LFQCD require close attention to the
LF singularities and zero-mode
contributions~\cite{t1,t2,t3,t4,t5,t6,t7,t8,t9,t10}. The
truncation of the light-front Fock-space calls for
the introduction of effective quark and gluon
degrees of freedom to overcome truncation effects,
e.g., see Refs.~\cite{TR1,TR2}. Introduction of
such effective degrees of freedom is what one
desires in seeking the dynamical connection
between canonical (or current) quarks and
effective (or constituent) quarks that Melosh
sought~\cite{TR3}, and Gell-Mann advocated as a
method for truncating QCD~\cite{TR4}.

The LF Hamiltonian formulation thus opens access
to QCD at the amplitude level and is poised to
become the foundation for a common treatment of
spectroscopy and the parton structure of hadrons
in a single covariant formalism, providing a
unifying connection between low-energy and
high-energy experimental data that so far remain
largely disconnected. 

\section{ APPLICATIONS OF THE LIGHT-FRONT FORMALISM }

\subsection{Structure of Hadrons}

Experiments that need a conceptually and
mathematically precise theoretical description of
hadrons at the amplitude level include
investigations of: the structure of nucleons and
mesons, heavy quark systems and exotics, hard
processes involving quark and gluon distributions
in hadrons, heavy ion collisions and many more.
For example, LFQCD will offer the opportunity for
an {\em ab initio} understanding of the
microscopic origins of the spin content of the
proton and how the intrinsic and spatial angular
momenta are distributed among the partonic
components in terms of the WFs. This is an
outstanding unsolved problem as experiments to
date have not yet found the largest components of
the proton spin. The components previously thought
to be the leading carriers, the quarks, have been
found to carry a small amount of the total spin.
Generalized parton distributions (GPDs) were
introduced to quantify each component of the spin
content, and the interface between GPDs and
experimental measurements in deeply virtual
Compton scattering (DVCS) has been discussed in
Ref.~\cite{JB1,JB2,JB3,JB4,JB5,JB6}. As another 
example, LFQCD will reproduce or predict the 
masses, quantum numbers and widths of the 
already familiar hadrons or yet-to-be observed 
exotics such as glueballs and hybrids. Some 
preliminary analyses can be found in 
Refs.~\cite{QQ1,QQ2,Hybrids1,Hybrids2}.

\subsection{QCD at High Temperature and Density}

There are major programs at accelerator facilities
such as GSI-SIS, CERN-LHC, and BNL-RHIC to
investigate the properties of a new state of
matter, the quark-gluon plasma, and other features
of the QCD phase diagram. In the early universe
temperatures were high, while net baryon densities
were low. In contrast, in compact stellar objects,
temperatures are low and the baryon density is
high. QCD describes both extremes. However,
reliable perturbative calculations can only be
performed at asymptotically large temperatures and
densities, where the running coupling constant of
QCD is small due to asymptotic freedom, and
lattice QCD provides information only at very low
chemical potential (baryon density). Thus, many
frontier questions remain to be answered. What is
the nature of the phase transitions? How does the
matter behave in the vicinity of the phase
boundaries? What are the observable signatures of
the transition in transient heavy-ion collisions?
LFQCD opens a new avenue for addressing these
issues. In recent years a general formalism to
directly compute the partition function in LF
quantization has been developed and numerical
methods are under development for evaluating this
partition function in
LFQCD~\cite{Elser:1996tq,Strauss:2008zx,Hiller:2007sc}. 
The goal is to establish a tool comparable in power to 
lattice QCD but extending the partition function to 
finite chemical potentials where experimental data 
are available.

\subsection{Nuclear Reactions}

There is a new appreciation that initial and
final-state interaction physics, which is not
intrinsic to the hadron or nuclear LFWFs, must be
addressed in order to understand phenomena such as
single-spin asymmetries, diffractive processes,
and nuclear shadowing (see the report
~\cite{Boer2011}). This motivates extending LFQCD
to the theory of reactions and to investigate
high-energy collisions of hadrons. Standard
scattering theory in Hamiltonian frameworks can
provide valuable guidance for developing a
LFQCD-based analysis of high-energy reactions.

\subsection{Intense Time-Dependent Fields}

High-intensity laser facilities offer prospects
for directly measuring previously unobserved
processes in QED, such as vacuum
birefringence~\cite{Heinzl:2006xc}, photon-photon
scattering~\cite{Lundstrom:2005za} and, still some
way in the future, Schwinger pair production.
Furthermore, `light-shining-through-walls'
experiments~\cite{Redondo:2010dp} can probe the
low energy frontier of particle physics and search
for beyond-standard-model
particles~\cite{Jaeckel:2010ni}. These
possibilities have led to great interest in the
properties of quantum field theories, in
particular QED, in background fields describing
intense light
sources~\cite{Heinzl:2008an,DiPiazza:2011tq}, and
some of the fundamental predictions of the theory
have been experimentally
verified~\cite{Bamber:1999zt}. 

Despite the basic theory behind `strong-field QED'
having been developed over 40 years ago, there
have remained until recent years several
theoretical ambiguities that can in part be
attributed to the use of the instant-form in a
theory which, because of the laser background,
naturally singles out light-like directions. Thus,
light-front quantization is a natural approach to
physics in intense laser fields. The use of the
front-form in strong-field
QED~\cite{Neville:1971uc} has provided answers to
several long standing questions, such as the
nature of the effective mass in a laser
pulse~\cite{Harvey:2012ie}, the pole structure of
the background-dressed
propagator~\cite{Ilderton:2012qe}, and the origins
of classical radiation reaction within
QED~\cite{Ilderton:2013dba}.

Combined with non-perturbative approaches such as
`time dependent basis light-front
quantization'~\cite{Zhao:2013cma, ZIMV}, which is
specifically targeted at time-dependent problems
in field theory, the front-form promises to
provide a better understanding of QED in external
fields. Such investigations will also provide
groundwork for understanding QCD physics in strong
magnetic fields at, for example, RHIC~\cite{QCD}.

\section{RELATIONSHIP WITH OTHER APPROACHES}

A solution of the LFQCD Hamiltonian eigenvalue
equation can utilize all available mathematical
methods of quantum mechanics and contribute to the
development of advanced computing techniques for
large quantum systems, including nuclei. For
example, in the Discretized Light Cone Quantization (DLCQ)~\cite{DQ1,DQ2,DQ3,DQ4,DQ5,DQ6}, 
periodic conditions are introduced such that momenta 
are discretized and the size of the Fock space is
limited without destroying Lorentz invariance.
Solving a quantum field theory is then reduced to
diagonalizing a large sparse Hermitian matrix. The
DLCQ method has been successfully used to obtain
the complete spectrum and LFWFs in numerous model
quantum field theories such as QCD with one or two
space dimensions for any number of flavors and
quark masses. An extension of this method to
supersymmetric theories, SDLCQ~\cite{Lunin:1999ib}, 
takes advantage of the fact that the LF Hamiltonian 
can be factorized as a product of raising and lowering 
ladder operators. SDLCQ has provided new insights into 
a number of supersymmetric theories including direct
numerical evidence~\cite{Hiller:2005vf} for a
supergravity/super-Yang--Mills duality conjectured
by Maldacena~\cite{Maldacena:1997re}.

One of the most interesting recent advances in
hadron physics has been the application to QCD of
a branch of string theory, Anti-de
Sitter/Conformal Field Theory
(AdS/CFT)~\cite{AdSCFT}. Although QCD is not a
conformally invariant field theory, one can use
the mathematical representation of the conformal
group in five-dimensional anti-de Sitter space to
construct an analytic first approximation to the
theory. The resulting model~\cite{H1,H2,H3,H4,H5,
H6,H7,H8,H9}, called AdS/QCD, gives accurate 
predictions for hadron spectroscopy and a description 
of the quark structure of mesons and baryons which 
has scale invariance and dimensional counting at 
short distances, together with color confinement 
at large distances. 

The dynamics in AdS space in five dimensions is
dual to a semiclassical approximation to
Hamiltonian theory in physical $3+1$ space-time
quantized at fixed light-front
time~\cite{deTeramond:2008ht}. Remarkably, there
is an exact correspondence between the
fifth-dimension coordinate of AdS space and a
specific impact variable $\zeta^2= b^2_\perp
x(1-x)$ which measures the physical separation of
the quark constituents within the hadron at fixed
light-cone time $\tau$ and is conjugate to the
invariant mass squared ${M^2_{q \bar q} }$. This
connection allows one to compute the analytic form
of the frame-independent simplified LFWFs for
mesons and baryons that encode hadron properties
and allow for the computation of exclusive
scattering amplitudes. 

The effective confining potential $U(\zeta^2)$  in this 
frame-independent ``light-front Schr\"odinger equation''
systematically incorporates the effects of higher quark 
and gluon Fock states.  The potential has a form of a 
harmonic oscillator potential if one requires that the 
chiral QCD action remains conformally
invariant~\cite{Brodsky:2013kpr}. The result is a 
nonperturbative relativistic light-front quantum 
mechanical wave equation which incorporates color 
confinement and other essential spectroscopic and 
dynamical features of hadron physics. 

These recent developments concerning AdS/CFT duality 
provide new insights about LFWFs which may form first 
approximations to the full solutions that one seeks in 
LFQCD, and be considered as a step in building a physically
motivated Fock-space basis set to diagonalize the LFQCD 
Hamiltonian, as in the ``basis light-front quantization'' 
(BLFQ) method~\cite{Vary:2009gt}. A complementary light-front
interpretation of the duality and holography is found 
in Ref.~\cite{Ehrenfest}.

\section{GOALS OF THE PROJECT}

The purpose of the LFQCD program is to bring
together experts in the field and attract new
contributors who will together take advantage of
the available theoretical and computational tools
and develop them further in order to provide
answers to the pertinent questions in an
accelerated fashion. The central issue is the 
rigorous description of hadrons, nuclei, and systems
thereof from first principles using QCD. We list 
the main goals of the required research.
\begin{enumerate}

\item 

Evaluation of masses and wave functions of hadrons
using the light-front Hamiltonian of QCD.

\item 

The analysis of hadronic and nuclear phenomenology
based on fundamental quark and gluon dynamics,
taking advantage of the connections between
quark-gluon and nuclear many-body methods.

\item 

Understanding of the properties of QCD at finite
temperatures and densities, which is relevant for
understanding the early universe as well as
compact stellar objects.

\item 

Developing predictions for tests at the new and
upgraded hadron experimental facilities -- JLAB,
LHC, J-PARC, GSI-FAIR.

\item 

Analyzing the physics of intense laser fields,
including a nonperturbative approach to
strong-field QED.

\item

Providing bottom-up fitness tests for model
theories as exemplified in the case of Standard
Model~\cite{V18}. 

\end{enumerate}

\noindent
To accomplish the nonperturbative analysis of QCD,
we need to:
\begin{enumerate}

\item 

Continue testing the LF Hamiltonian approach in
simple theories in order to improve our
understanding of its peculiarities and treacherous
points vis \`{a} vis manifestly-covariant
quantization methods~\cite{t1,t2,t3,t4,t5,t6,t7,t8,t9,t10}.
This will include work on theories such as Yukawa
theory~\cite{kms2012,Y-1,Y0,Y1,Y2,Y3,Y4} and 
QED~\cite{QED0,QED1a,QED1b,QED2,QED3,QED4} and on 
theories with unbroken supersymmetry, in order to 
understand the strengths and limitations of different 
methods. Much progress has already been made along 
these lines.

\item 

Construct most symmetry-preserving regularization 
and renormalization schemes for light-front QCD, to
take practical advantage of the Pauli--Villars-based 
method of the St. Petersburg group~\cite{SP1,SP2}, 
G{\l}azek--Wilson similarity renormalization-group 
procedure for Hamiltonians~\cite{GW1,GW2,GW3} 
(Wilsonian concept of coupling constant 
renormalization~\cite{KGW1} is made available 
in its LF version in~\cite{KGWSDGRJP}),
Mathiot--Grang\'{e} test functions~\cite{Mathiot-Grange}, 
Karmanov-Mathiot-Smirnov~\cite{kms2012} 
realization of the sector-dependent
renormalization~\cite{sdr1,sdr2,sdr3,sdr4},
and determine how to
incorporate symmetry breaking in light-front
quantization~\cite{ss1,ss2,ss3,ss4,ss5,ss6}; this is
likely to require an analysis of zero modes and
in-hadron condensates~\cite{V13,V14,V15,V17}.

\item 

Develop computer codes which implement the
regularization and renormalization
schemes.\footnote{An example of a related
discussion is available at
www.fuw.edu.pl/$\sim$lfqcd/inmemoriam/?part=20.}
Provide a platform-independent, well-documented
core of routines that allow investigators to
implement different numerical approximations to
field-theoretic eigenvalue problems, including the
light-front coupled-cluster
method~\cite{Chabysheva:2011ed}. Consider various
quadrature schemes and basis sets, including DLCQ,
finite elements, function
expansions~\cite{ref.VanIersel}, and the complete
set of orthonormal wave functions obtained from
AdS/QCD~\cite{bf1,bf2,bf3}. This will build on the
Lanczos-based MPI code developed for
nonrelativistic nuclear physics~\cite{Sternberg:2008, Maris:2010iccs}
applications and
similar codes for Yukawa theory and
lower-dimensional supersymmetric Yang--Mills
theories.

\item

Address the problem of computing theoretical
bounds on truncation errors and other ambiguities
introduced by various simplifying assumptions,
particularly for energy scales where QCD is
strongly coupled. Understand the role of
renormalization group methods~\cite{inftynumber,
couplings}, asymptotic freedom~\cite{PerryAF, effgluons} 
and spectral properties of
$P^+$ in quantifying theoretical errors, as one
could do in the case of model LF lattice
dynamics~\cite{DalleyLattice} or in model studies
of mathematical accuracy of the similarity 
renormalization group procedure for Hamiltonians 
in Refs.~\cite{GlazekMlynikOptimization,
GlazekMlynikBenchmark}. Such studies of
theoretical accuracy are necessary for
understanding and differentiating between inputs
characterizing various approaches when estimating
their predictive power and capability of
falsifying theories.

\item 

Solve eigenvalue problems for hadronic masses and 
wave functions, cf.~\cite{bf3}. Use these wave functions 
to compute form factors, GPDs, scattering amplitudes, 
and decay rates. Compare with perturbation theory, 
lattice QCD, and model calculations, using insights 
from AdS/QCD, where possible. Study the transition 
to nuclear degrees of freedom, beginning with light 
nuclei.

\item 

Classify the spectrum with respect to total
angular momentum. In equal-time quantization, the
three generators of rotations are kinematic, and
the analysis of total angular momentum is
relatively simple. In light-front quantization,
only the generator of rotations around the
$z$-axis is kinematic; the other two, of rotations
about the axes $x$ and $y$, are dynamical. To
solve the angular momentum classification problem,
the eigenstates and spectra of the sum of squares
of these generators must be
constructed~\cite{LeutAnnPhys78, Algebra}. This is
the price to pay for having more kinematical
generators than in equal-time quantization, where
all three boosts are dynamical. In light-front
quantization, the boost along $z$ is kinematic,
and this greatly simplifies the calculation of
matrix elements that involve boosts, such as the
ones needed to calculate form factors. The
relation to covariant Bethe-Salpeter approaches
projected on the
LF~\cite{vak1980,BS1,BS2,BS3,BS4,BS5,BS6,BS7} may
help in understanding the angular momentum issue
and its relationship to the Fock-space truncation
of the LF Hamiltonian. Model-independent
constraints from the general angular
condition~\cite{vak1982a,vak1982b,Carlson-Ji},
which must be satisfied by the LF helicity
amplitudes, should also be explored. The
contribution from the zero mode appears necessary
for the hadron form factors~\cite{kms2007} to
satisfy angular momentum conservation, as
expressed by the angular
condition~\cite{AC-spin1,AC-spin12}. 
The relation to light-front quantum mechanics,
where it is possible to exactly realize full
rotational covariance and construct explicit
representations of the dynamical rotation
generators, should also be explored.

\item 

Explore the AdS$_5$/QCD correspondence and
light-front holography~\cite{H1,H2,H3,H4,H5,H6,H7,H8,H9}. 
The approximate duality in the limit of massless 
quarks motivates few-body analyses of meson and 
baryon spectra based on a one-dimensional 
light-front Schr\"odinger equation in terms of 
the modified transverse coordinate $\zeta$. Models
that extend the approach to massive quarks have
been proposed, but a more fundamental
understanding within QCD is needed. The nonzero
quark masses introduce a non-trivial dependence on
the longitudinal momentum, and thereby highlight
the need to understand the representation of
rotational symmetry within the formalism.
Exploring AdS$_5$/QCD wave functions as part of a
physically motivated Fock-space basis set to
diagonalize the LFQCD Hamiltonian should shed
light on both issues. The complementary Ehrenfest
interpretation can be used to introduce effective
degrees of freedom such as diquarks in
baryons~\cite{Ehrenfest}.

\item 

Develop numerical methods/computer codes to
directly evaluate the partition function ({\em
viz}.\ thermodynamic potential) as the basic
thermodynamic quantity. Compare to lattice QCD,
where applicable, and focus on a finite chemical
potential, where reliable lattice QCD results are
presently available only at very small (net) quark
densities. There is also an opportunity for use of
LF AdS/QCD to explore non-equilibrium phenomena
such as transport properties during the very early
state of a heavy ion collision. LF AdS/QCD opens
the possibility to investigate hadron formation in
such a non-equilibrated strongly coupled
quark-gluon plasma.

\item

Develop a LF approach to the neutrino oscillation
experiments that are possible at Fermilab and 
elsewhere, with the goal of reducing the energy 
spread of the neutrino-generating hadronic sources, 
so that the three-energy-slits interference picture 
(assuming there exist only three neutrinos) of the
oscillation pattern~\cite{neutrinos1,neutrinos2} 
can be resolved and the front form of Hamiltonian
dynamics utilized in providing the foundation for
qualitatively new (treating the vacuum differently
than it is treated in the instant form of dynamics) 
studies of neutrino mass generation mechanisms. 

\item

Take advantage of the possibility that, if the 
renormalization group procedure for effective 
particles (RGPEP)~\cite{RGPEP1,RGPEP2} does allow 
one to study intrinsic charm, bottom, and glue in a 
renormalized and convergent LF Fock-space expansion, 
one might consider a host of new experimental studies 
of production processes using the intrinsic components 
that are not included in the calculations based on 
gluon and quark splitting functions. 

\end{enumerate}

\section{ Conclusion }

As a theory and foundation for the phenomenology of processes
involving hadrons, QCD faces challenges that by no means are
resolved, neither directly nor at the current conceptual level
of attempts to improve the standard model and seek a unified
theory beyond it.  A hadron eigenstate of the LFQCD Hamiltonian,
calculated with modern tools of massive computing, can provide
previously unavailable capabilities for in-depth exploration
of the structure of the Fock-space wave functions.  The
discovery potential hidden in LFQCD for understanding basic
theoretical issues in particle physics is as great as the
utility of this approach as a tool, deeply rooted in theory,
for the phenomenology of strong interactions.



\end{document}